\documentclass[journal]{IEEEtran}
\ifCLASSINFOpdf
\else
\fi

\usepackage{booktabs} 
\usepackage{subfig}
\usepackage{graphicx}
\usepackage{algorithm}
\usepackage{algpseudocode}%
\usepackage{xcolor}
\usepackage{amsmath}
\newcommand{\etal}{\textit{et al}. }
\usepackage{mathtools}
\usepackage{multirow}
\usepackage{graphics}
\usepackage{enumitem}
\usepackage{CJKutf8}
\usepackage{array}
\usepackage{diagbox}
\usepackage{hyperref}
\usepackage[utf8]{inputenc}
\usepackage{ragged2e}
\usepackage{amsmath,amssymb}

\usepackage{xcolor,cite,etoolbox}
\definecolor{mypurple}{rgb}{0.4392, 0.1882, 0.6275}
\definecolor{mygreen}{rgb}{0, 0.6902, 0.3137}
\makeatletter 
\pretocmd\@bibitem{\csname keycolor#1\endcsname}{}{\fail}
\newcommand\citecolor[1]{\@namedef{keycolor#1}{}}
\makeatother
\begin{document}
%
\title{PR-PL: A Novel Transfer Learning Framework with \\Prototypical Representation based Pairwise Learning \\for EEG-Based Emotion Recognition}
%
%
%
\author{\IEEEauthorblockN{Rushuang Zhou\textsuperscript{a,b,1,}\IEEEauthorrefmark{1},
Zhiguo Zhang\textsuperscript{a,b,c,d,e,1,}\IEEEauthorrefmark{2},
Hong Fu\textsuperscript{f,}\IEEEauthorrefmark{3},
Li Zhang\textsuperscript{a,b,}\IEEEauthorrefmark{4},
Linling Li\textsuperscript{a,b,}\IEEEauthorrefmark{5},\\
Gan Huang\textsuperscript{a,b,}\IEEEauthorrefmark{6},
Yining Dong\textsuperscript{g,}\IEEEauthorrefmark{7},
Fali Li\textsuperscript{h,i,}\IEEEauthorrefmark{8},
Xin Yang\textsuperscript{a,b,}\IEEEauthorrefmark{9} and Zhen Liang\textsuperscript{a,b,}\IEEEauthorrefmark{10}}\\
\medskip
\IEEEauthorblockA{\small{{\textsuperscript{a}School of Biomedical Engineering, Health Science Center, Shenzhen University, Shenzhen, China\\
\textsuperscript{b}Guangdong Provincial Key Laboratory of Biomedical Measurements and Ultrasound Imaging, Shenzhen, China\\
\textsuperscript{c}chool of Computer Science and Technology, Harbin Insti-
tute of Technology, Shenzhen, China\\
\textsuperscript{d}Marshall Laboratory of Biomedical Engineering, Shenzhen, China\\
\textsuperscript{e}Peng Cheng Laboratory, Shenzhen, China\\
\textsuperscript{f}Department of Mathematics and Information Technology, The Education University of Hong Kong, Hong Kong\\
\textsuperscript{g}School of Data Science, City University of Hong Kong, Hong Kong\\
\textsuperscript{h}The Clinical Hospital of Chengdu Brain Science Institute, MOE Key Lab for Neuroinformation,\\ University of Electronic Science and Technology of China, China\\
\textsuperscript{i}School of Life Science and Technology, Center for Information in Medicine,\\ University of Electronic Science and Technology of China, China\\
\medskip
Email: \IEEEauthorrefmark{1}2018222087@szu.edu.cn,
\IEEEauthorrefmark{2}zgzhang@szu.edu.cn,
\IEEEauthorrefmark{3}hfu@eduhk.hk,
\IEEEauthorrefmark{4}lzhang@szu.edu.cn,
\IEEEauthorrefmark{5}lilinling@szu.edu.cn,\\
\IEEEauthorrefmark{6}huanggan@szu.edu.cn,
\IEEEauthorrefmark{7}yinidong@cityu.edu.hk,
\IEEEauthorrefmark{8}fali.li@uestc.edu.cn,
\IEEEauthorrefmark{9}yangxinknow@gmail.com,
\IEEEauthorrefmark{10}janezliang@szu.edu.cn}}}}

\maketitle

\begin{abstract}
Affective brain-computer interfaces based on electroencephalography (EEG) is an important branch in the field of affective computing. However, individual differences and noisy labels seriously limit the effectiveness and generalizability of EEG-based emotion recognition models. In this paper, we propose a novel transfer learning framework with Prototypical Representation based Pairwise Learning (PR-PL) to learn discriminative and generalized prototypical representations for emotion revealing across individuals and formulate emotion recognition as pairwise learning for alleviating the reliance on precise label information. More specifically, a prototypical learning-based adversarial discriminative domain adaptation method is developed to encode the inherent emotion-related semantic structure of EEG data, while pairwise learning with an adaptive pseudo-labeling method is developed to achieve a reliable and stable model learning with noisy labels. Through domain adaptation, feature representations of source and target domains are aligned on a shared feature space, while the feature separability of both source and target domains is also considered. The characterized prototypical representations are evident with a high feature concentration within one single emotion category and a high feature separability across different emotion categories. Extensive experiments are conducted on two benchmark databases under four cross-validation evaluation protocols (cross-subject cross-session, cross-subject within-session, within-subject cross-session, and within-subject within-session). The experimental results demonstrate the superiority of the proposed PR-PL against the state-of-the-arts under all four evaluation protocols, which shows the effectiveness and generalizability of PR-PL in dealing with the ambiguity of EEG responses in affective studies. The source code is available at \textit{https://github.com/KAZABANA/PR-PL}.
\end{abstract}


\footnotetext[1]{\hspace{1mm}Equal contributions.}

%
\IEEEpeerreviewmaketitle

\section{Introduction}
\label{sec:introduction}
\IEEEPARstart{A}{ffective} computing is a fast growing interdisciplinary research field and is attracting researchers' attention from different areas including computer science, neuroscience, psychology, and signal processing \cite{siddharth2019utilizing}. Recently, electroencephalography (EEG) based emotion recognition has become an increasingly important topic for affective computing and human sentiment analysis \cite{hu2019ten,hu2020video}. A proper design of EEG-based emotion recognition models is helpful for facilitating the data processing, benefiting discriminant feature characterization, and lightening the model performance. Currently, there exist two main critical issues in EEG-based emotion recognition. One is \textbf{individual differences}: how to build a generalized affective computing model which could tolerate the remarkable individual differences in the simultaneously collected EEG signals; and another is \textbf{noisy label learning}: how to train a reliable and stable affective computing model which is less reliant on the subjective feedback.

In recent years, more and more researchers have focused on applying transfer learning methods to alleviate the individual differences in EEG signals \cite{TLBCI2016,li2019multisource,li2019regional,cui2020eeg,zhong2020eeg,gu2021eeg} and improve feature invariant representation \cite{ozdenizci2019adversarial,ozdenizci2020learning,bethge2022domain}. Considering the individuals with and without labels (termed as \textbf{source domain} and \textbf{target domain}),  transfer learning tries to minimize the distribution difference between the source and target domains by approximately satisfying the assumption of independent and identical distribution and can consequently realize a higher recognition performance on the target domain. Through a domain-shifting strategy, the invariant feature representations across different domains are learned and the relationships among the learned features, data distribution, and labels are explored. For example, Li \etal \cite{li2019multisource} proposed a multisource transfer learning method with two transfer learning stages. In the first stage, appropriate samples were selected from the existing source domain. In the second stage, a style transfer mapping was implemented to alleviate the differences between the selected source samples and the unknown target samples. The results showed the proposed transfer method outperformed the non-transfer method with an improvement of 12.72\% on the public SEED database \cite{zheng2015investigating} with a three-class classification problem (negative, neutral, and positive). Inspired by neuroscience findings that different emotions would lead to different brain reactions, Li \etal \cite{li2019regional} proposed a novel R2G-STNN network to integrate the EEG spatial-temporal dynamics at the local and global brain areas and realize an efficient emotion recognition performance together with a domain shift learning. More details about current EEG-based emotion recognition models with the transfer learning algorithms are presented in Section \ref{sec:relatedWork}. 

For video-evoking EEG-emotion experiments, subjects may not always be able to accurately react to the intended emotions, and at the same time may not be able to accurately describe and feedback on their emotional changes. This would bring label noise to the emotional information annotation of EEG samples and further lead to a negative impact on the model performance \cite{Jia2010Noise}. To tackle this issue, Zhong \etal \cite{zhong2020eeg} developed an emotion-aware distribution learning method (RGNN), in which they blurred the label information by changing the one-hot label representation $\left(1,0,0\right)$ to $\left(1-\frac{2\epsilon}{3},\frac{2\epsilon}{3},0\right)$ and trained the model to be less sensitive label noise. However, the model performance would greatly rely on the selection of $\epsilon$ value, and an optimal $\epsilon$ value selection could be different for different databases and different individuals. Current EEG-based emotion recognition models are mainly based on \textbf{pointwise learning}, which heavily relies on precisely labeled data. Contrarily, \textbf{pairwise learning} makes it possible to model the relative associations between pairs of instances and to efficiently encode the proximity among samples with less reliance on labeling. Thus, pointwise learning has achieved tremendous success in a number of real-world applications \cite{bao2018classification,bao2020similaritybased,hsu2020deep,zhuang2020learning}.

To further improve the effectiveness and generalizability of EEG-based emotion recognition models and eliminate the negative effects from individual differences and label noises, in this paper, we formulate the emotion recognition tasks as a pairwise learning problem and propose a novel transfer learning framework with prototypical representation based pairwise learning (which is termed as \textbf{PR-PL} below). Here, we model the relative relationship between pairs of EEG samples in terms of prototypical representations, which is advantageous to pointwise learning when the labeling task is difficult and even the provided labels are wrong labels\cite{yao2018representation}. The major novelties of the proposed PR-PL are summarized as follows. (1) We formulate emotion recognition as pairwise learning to replace the classifier and greatly alleviate the label dependence on emotion labels. The pairwise learning provides us an alternative way to measure whether two EEG signals belong to the same emotion category without the reliance on the precise labeling information. The extensive experimental results on two well-known emotional databases (SEED \cite{zheng2015investigating} and SEED-IV \cite{zheng2018emotionmeter}) prove the proposed PR-PL is 
a more accurate model than the state-of-the-arts for solving the EEG-based emotion recognition tasks under different application environments (cross-subject cross-session, cross-subject single-session, within-subject cross-session, and within-subject single-session). (2) We propose a novel prototypical learning-based adversarial discriminative domain adaptation method to explore latent variables of emotion categories, encode the semantic structure of EEG data, and learn subject-generalized prototypical representations for emotion revealing across individuals. The characterized prototypical representations show a high feature concentration within one single emotion category and a high feature separability across different emotion categories. (3) Different from the existing transfer learning methods that only focus on feature separability in the source domain, we consider the feature separability of both source and target domains through the end-to-end domain adversarial training to further enhance the model effectiveness and generalizability.

\section{Related Work}
\label{sec:relatedWork}
The existing EEG-based emotion recognition models with the transfer learning algorithms can be generally categorized into two types. 

{\textbf{(a) Non-deep transfer learning models.}} Pan \etal \cite{TCA2010} proposed a transfer component analysis (TCA) algorithm to reduce the marginal distribution difference between the source and target domains, in which the transfer information was learned in a reproducing kernel Hilbert space through maximizing mean discrepancy. Zheng and Lu\cite{Zhengpersonal2016} \textcolor{black}{introduced} two types of subject-to-subject transfer methods to deal with the challenge of the individual differences in EEG signal processing. One was to explore a shared common feature space underlying source and target domains using TCA and kernel principal analysis (KPCA), and another was to construct multiple personalized classifiers on the source domain and map the classifier parameters to the target domain using transductive parameter transfer (TPT). These non-deep transfer learning strategies show the possibility to bridge the discrepancy across two domains with improved performance on the target domain. However, due to the small capacity and low complexity, the model accuracy and stability are still limited, which fails to satisfy the requirements of affective brain-computer interfaces (aBCI) in practical applications. 

{\textbf{(b) Deep transfer learning models.}} Most of the existing affective models are based on deep transfer learning methods built with domain-adversarial neural network (DANN) proposed in \cite{ganin2016domain}. The main idea of DANN is to find a shared feature representation for source and target domains with indistinguishable distribution differences and also maintain the predictive ability of the estimated features on the source samples for a specific classification task. Li \etal \cite {He2018DAN} was the first to introduce DANN in aBCI. Benefiting from the powerful feature representation ability of deep networks and the high efficiency of adversarial learning in distributed adaptation, the results showed that DANN based aBCI system was superior to other methods. The following aBCI systems could be considered as a series of DANN-based models, which generally start from two directions to improve the DANN performance in solving EEG-based emotion recognition tasks. 
\begin{itemize}
\item \textbf{Incorporating the prior knowledge of neuroscience and brain anatomy with DANN.} Inspired by the neuroscience findings of the asymmetry property of the left and right hemispheres in emotional responses, Yang \etal \cite{li2018bi} proposed a bi-hemisphere domain adversarial neural network (BiDANN), in which a global and two local domain discriminators were designed to learn discriminant features from each cerebral hemisphere related to emotion perception and also improve the feature stability to the variation of different domains. The experiments on the SEED database demonstrated that BiDANN achieved higher emotion recognition performance than DANN. Considering the emotional responses from different brain regions would be varied, Yang \etal \cite{li2019regional} proposed an R2G-STNN (regional to global-spatial-temporal neural network) to integrate the spatial-temporal information from local and global brain regions under importance guidance and characterize hierarchical feature representations. Similarly, under an assumption that not all EEG channels are equally important in emotion recognition tasks, Du \etal \cite{Du2020} integrated attention mechanism, long-short-term memory (LSTM), and DANN to propose an attention-based LSTM with domain discriminator (ATDD-LSTM) and characterize the nonlinear relations among different EEG channels in a data-driven approach and optimally select informative emotion-related channels.
\item \textbf{Incorporating the probability distribution with DANN.} To deal with the training instability of DANN, Luo \etal \cite{2018WGAN} introduced WGAN-GP (Wasserstein generative adversarial network with gradient penalty) to narrow down the distance between the marginal probability distributions of different subjects. The results showed the model stability was improved and a better cross-subject EEG-based emotion recognition was achieved. However, current DANN-based models only consider the marginal distribution differences but ignore the joint distribution differences of different domains. To address this problem, Li \etal \cite{LiJDA2020} introduced a joint domain adaptation network (JLNN), where the joint distribution adaptation (JDA) method was incorporated with a unified framework of task-invariant features (MDA) and task-specific features (CDA). 
\end{itemize}

Although the above models have achieved higher accuracies compared to the original model with DANN in emotion recognition tasks, there still exist three major technical challenges. First, the learned feature representation is susceptible to noise interference from both source and target domains and would further affect the model generalizability \cite{2010Factorized,Bous2016DSN}. Second, the existing models only focus on the feature separability in the source domain but ignore the feature separability in the target domain. The current DANN-based models mainly concern the emotion classification loss in the source domain, which would lead to the over-fitting of source domain data and the decrease of classification ability on target domain data. Third, the existing algorithms largely rely on a large amount of labeled source domain data. However, in practical EEG applications, it is difficult to collect accurate labels for each single EEG trial.

\begin{figure*}
\begin{center}
\includegraphics[width=1\textwidth]{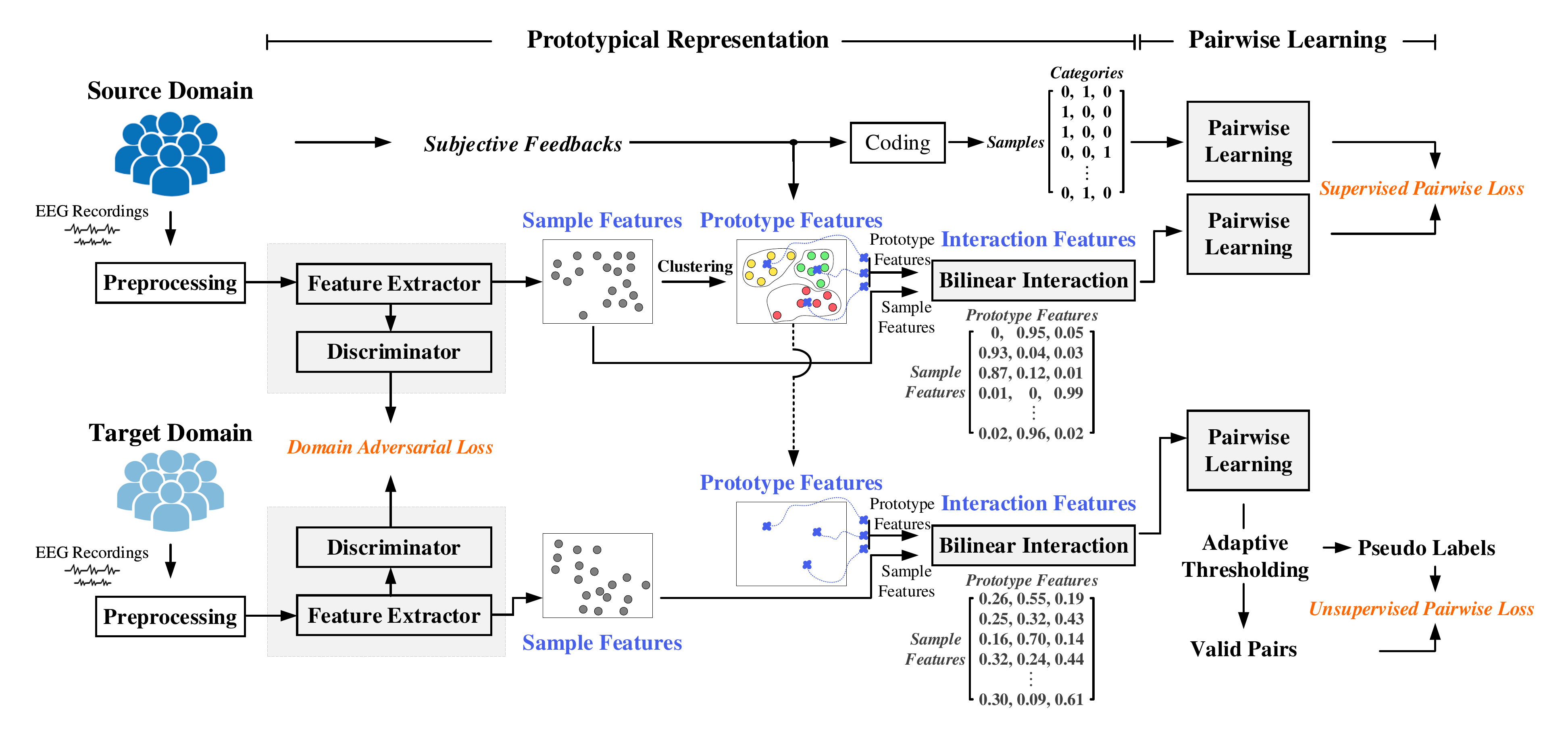}
\end{center}
\caption{The proposed PR-PL framework.}
\label{fig:PR-PLmodel}
\end{figure*}

\begin{table}
\centering
\caption{Frequently used notations and descriptions.}
\label{tab:frequentNotations}
\begin{tabular}{cc}
\toprule
 Notation             & Description                                   \\
\midrule
 $\mathbb{S}    \backslash \mathbb{T} $                & source\textbackslash{}target domain   \\
 $x^s \backslash x^t$ & source\textbackslash{}target feature   \\
 $y^s \backslash y^t$ & source\textbackslash{}target label      \\
 $N^s \backslash N^t$ & number of source\textbackslash{}target samples\\
 $D^s \backslash D^t$ & the source\textbackslash{}target dataset\\
 $f\left(\cdot\right)$& sample feature extractor                      \\
 $d\left(\cdot\right)$& domain discriminator                          \\
 $h\left(\cdot\right)$& bi-linear operation                           \\
 $\theta_f           $& the parameter of the feature extractor $f\left(\cdot\right)$\\
 $\theta_d           $& the parameter of the discriminator $d\left(\cdot\right)$\\
 $S                  $& the parameter of the bi-linear operation $h\left(\cdot\right)$\\
 $\mu$                & centroid                                      \\
 $l$                  & predict label                                 \\
\bottomrule
\end{tabular}
\end{table}

\section{Methodology}
\label{sec:methodology}
Suppose the EEG trials in the source domain $\mathbb{S}$ and target domain $\mathbb{T}$ are given as ${D}_{s}=\{X_s,Y_s\}$ and ${D}_{t}=\{X_t,Y_t\}$, where ${\{X_s,Y_s\}=\left\{\left(x_{i}^s,y_{i}^s\right)\right\}_{i=0}^{N_s}}$ and ${\{X_t,Y_t\}=\left\{\left(x_{i}^t,y_{i}^t\right)\right\}_{i=0}^{N_t}}$. Here, $X_{i}^s$ and $X_{i}^t$ are EEG samples, and $Y_{i}^s$ and $Y_{i}^t$ are the corresponding emotion labels. To make the narrative clearer, the frequently used notations are summarized in Table \ref{tab:frequentNotations}. \textcolor{black}{As shown in Fig. \ref{fig:PR-PLmodel}, the proposed PR-PL includes three losses (domain adversarial loss, pairwise learning loss on source domain, and pairwise learning loss on target domain) and two main parts (prototypical representation and pairwise learning). In the prototypical representation, three types of features are defined. The characterized features $f(X_s)$ or $f(X_t)$ from the EEG samples are termed as \textbf{sample features}, which are forced to be as indistinguishable as possible from the source and target domains. Under an assumption that any emotion could be represented by a prototype via prototypical learning, the \textbf{prototype features} of each emotion category are learned based on the sample features $f(X_s)$ and $Y_s$ from the source domain. The interaction relationships between the sample features and prototype features are measured and the \textbf{interaction features} are characterized which will be used in the following pairwise learning. In the pairwise learning, the pair relationships on both source and target domains are explored. As the information about $Y_t$ are unknown during model training, an adaptive thresholding method is developed for valid pair selection and pseudo label generation.} 

\subsection{Sample feature extraction}
\label{sec:sampleFea}
To make both source and target data satisfy the assumption of independent and identical distribution and obey the same distribution, we characterize the sample features based on a domain adversarial training introduced in DANN. Here, the distribution difference between the source domain and the target domain is alleviated and the sample features with domain invariant properties are characterized. Through this process, the individual differences in EEG signals could be alleviated and the generalization ability of the models could be improved \cite{li2018bi,2018WGAN,li2019regional,Du2020}. Specifically, the domain difference between the source domain sample feature $f(X_s)$ and the target domain sample feature $f(X_t)$ are minimized by adopting domain adversarial training. Here, $f\left(\cdot\right)$ is the designed feature extractor for extracting the sample features from EEG signals. A discriminator network $d\left(\cdot\right)$ with the parameter $\theta_d$ is introduced to distinguish whether the characterized sample features ($f(X_s)$ or $f(X_t)$) come from the source domain ($\mathbb{S}$) or the target domain ($\mathbb{T}$). Its loss function is a standard two-category cross-entropy loss function, given as
\begin{equation}
    \label{Eq:DANNdiscriminator}
        \begin{aligned}
        \mathcal{L}_{d i s c}\left(\theta_{f}, \theta_{d}\right)=&-\sum_{i=0}^{N_{s}} \log d\left(f\left(x_{i}^{s}\right)\right)-
        \sum_{i=0}^{N_{t}} \log \left(1-d\left(f\left(x_{i}^{t}\right)\right)\right).
\end{aligned}
\end{equation}

In the training process, we adopt the end-to-end training method \cite{ganin2016domain}, and implement the domain adversarial training by introducing a gradient reversal layer. The feature extractor $f(\cdot)$ maximizes the classification ability to enhance emotion recognition performance and at the same time, the discriminator $d(\cdot)$ minimizes the domain discrimination to reduce the distribution difference between the source domain and target domain. The final domain adversarial training objective function is defined as

\begin{equation}
    \label{Eq:DANNfull}
        \min _{\theta_{f}}\max _{\theta_{d}} \mathcal{L}_{\text{classifier}}^{s}(\theta_{f})-\lambda \mathcal{L}_{\text{disc}}\left(\theta_{f}, \theta_{d}\right),
\end{equation}
which is termed as \textbf{domain adversarial loss} in Fig. \ref{fig:PR-PLmodel}. Here, $\mathcal{L}_{\text{classifier}}^{s}(\theta_{f})$ is the classification loss to measure the classification ability in source domain. In this paper, $\mathcal{L}_{\text{classifier}}^{s}(\theta_{f})$ will be realized by pairwise learning on source and target domains introduced in Section \ref{sec:pairwiseLearnsource} and \ref{sec:pairwiseLearnTarget} below. $\mathcal{L}_{\text{disc}}(\theta_{f},\theta_{d})$ is the adversarial loss for the discriminator to be trained to distinguish the sample features characterized from source and target domains. $\theta_{f}$ and $\theta_{d}$ are the parameters of $f\left(\cdot\right)$ and $d\left(\cdot\right)$. $\lambda$ is a balanced hyperparameter for ensuring the stability of domain adversarial, which is given by a exponential growth method as
\color{black}
\begin{equation}
    \label{Eq:DANNlambda}
        \lambda=\frac{2}{1-\exp (-p)}-1.
\end{equation}
Here, $p$ is a factor related to the training round, given by a ratio of the current training round to the maximum training round.

\color{black}
\subsection{Prototype feature extraction}
\label{sec:prototypeFea}
We assume that there exists a prototype for each emotion category. Through prototypical learning, the prototype features are learned to indicate the representation property of every single emotion category. Based on the sample features extracted from different subjects under different emotions, we could consider these sample features are distributed around the prototype features. In other words, for each emotion category, the prototype features could be considered the "center of mass" of all the sample features. From the perspective of a probability distribution, the prototype feature of an emotion category can be regarded as the mean value of the sample feature distribution of the emotion, and the variance of the distribution is caused by the non-stationary EEG, including but not limited to individual differences. Assume that the sample features under an emotion category $c$ obey the Gaussian distribution $N\left( \mu_{c},\sigma_{c}^{2}\right)$. The prototype features of the emotion category $c$ could be calculated as the mean vector $\mu_{c}$ of the distribution. For the source domain data {$\{{X}_{s},{Y}_{s}\}=\left\{\left(x_{i},y_{i}\right)\right\}_{i=0}^{N_s}$}, the corresponding sample features are characterized by the feature extractor $f\left(\cdot\right)$, given as $f(x_{i})$ (defined in Section \ref{sec:sampleFea}). The prototype feature vector of the emotion category $c$ can be calculated by averaging all the sample features that belong to this category, given as

 \begin{equation}
    \label{Eq:centroid}
    \mu_{c}=\frac{1}{\left|{X}_{s}^{c}\right|} \sum_{x_{i}^{s} \in {X}_{s}^{c}} f\left(x_{i}^{s}\right),
\end{equation}
where ${X}_{s}^{c}=\left\{\left(x_{i}^{s},y_{i}^{s}=c\right)\right\}_{i=0}^{N}$ are a collection of source domain data belonging to the emotion category $c$. $\left|{X}_{s}^{c}\right|$ are the corresponding sample size in this emotion category. In other words, $\mu_{c}$ could be expressed as the centroid of the sample features of ${X}_{s}^{c}$. The mean value calculation is a widely used, simple, and effective noise reduction strategy, which can make the prototype features stronger than the sample features and help to alleviate the problem of the traditional DANN network being susceptible to related noise interference \cite{Pinheiro2018}. It is worth noting that since the calculation of the prototype features needs to use emotional label information, we only use the source domain data here for prototype feature extraction.

\subsection{Interaction feature extraction}
\label{sec:interactionFea}
The traditional DANN extracted domain-invariant shared feature representations could be easily contaminated by the shared and related noises in source and target domains. In this paper, we \textcolor{black}{introduce} a bilinear interaction to measure the interaction relationships between the sample features and prototype features and extract the interaction features for the following pairwise learning. For a given $d$-dimensional sample feature $f(x_i)$, its interaction relationship to a certain prototype feature $\mu_(c)$ can be measured by a bilinear transformation $h(\cdot)$, defined as
 \begin{equation}
    \label{Eq:bilinear}
        h\left(f\left(x_{i}\right), \mu_{c}\right)=f\left(x_{i}\right)^{T}S \mu_{c},
\end{equation}
where $S\in{R^{d\times d}}$ is a bilinear transformation matrix with trainable parameters and is not restricted by symmetry or positive definiteness. Suppose there are total $n$ emotion categories, the interaction measurement between a certain sample feature $f\left(x_{i}\right)^{T}$ and different prototype features can be represented as
\begin{equation}
    \label{Eq:interactionFea}
        l_{i}=\text{softmax}\left(\left[h\left(f\left(x_{i}^{s}\right), \mu_{1}\right)          ,...,h\left(f\left(x_{i}^{s}\right), \mu_{n}\right)\right]\right),
\end{equation}
where $\{\mu_i\}_{i=1}^n$ are all the prototype features. An introduction of the activation function (softmax) here is to add nonlinear advantages to the interaction measurement, enhance the feature representation ability, and at the same time allow the feature vector $l$ to have category prediction capabilities.

\subsection{Pairwise learning on source domain}
\label{sec:pairwiseLearnsource}

Traditional pointwise learning algorithms often regard the feature vector $l$ as the predicted label of the sample feature $x_{i}$ and use the cross-entropy loss function to match the label $y_{i}$ of $x_{i}$ based on supervised training. Then, if the sample features of one EEG trial match the prototype features of $j$th emotion category the most, the EEG trial will be assigned to the $j$th emotion category. However, this type of pointwise learning only focuses on the relationship between sample features and prototype features and ignores the relationship between different sample features. For example, the sample features belonging to different emotion categories should be separated from each other, and the sample features belonging to the same emotion category should be gathered together. To tackle this issue, we introduce pairwise learning to capture the inherent relationship of samples. 

For the source domain data $\{X_s,Y_s\}=\{x_{i}^{s},y_{i}^{s}\}_{i=1}^{N_s}$, the corresponding loss function for pairwise learning is defined as
\begin{equation}
    \label{Eq:pairwiseSource}
        \mathcal{L}_{\text {class }}(\theta)
        =\sum_{i, j} L\left(r_{i j}^{s}, g\left({x}_{i}^{s}, {x}_{j}^{s};\mathbf{\theta}\right)\right),
\end{equation}
where $g\left({x}_{i}^{s}, {x}_{j}^{s} ; \mathbf{\theta}\right)$ is the similarity measurement of the samples $x_i^s$ and $x_j^s$, with the parameter of $\theta$. According to the assumption of pairwise learning, if ${y}_{i}^{s}={y}_{j}^{s}$, then $r_{i j}^{s}=1$; otherwise $r_{i j}^{s}=0$. The loss function $L\left(\cdot\right)$ is a difference calculation of $r_{i j}$ and $g\left({x}_{i}^{s}, {x}_{j}^{s} ; \mathbf{\theta}\right)$, given by a two-category cross-entropy loss as
 \begin{equation}
    \label{Eq:pairwiseSourceL}
    \begin{aligned}
    &L\left(r_{i j}^{s}, g\left(x_{i}^{s}, x_{j}^{s} ; \theta\right)\right)=
    -r_{i j}^{s} \log \left(g\left(x_{i}^{s},x_{j}^{s} ; \theta\right)\right)\\
    &-\left(1-r_{i j}^{s}\right) \log \left(1-g\left(x_{i}^{s},x_{j}^{s};\theta\right)\right).
    \end{aligned}
\end{equation}
In the training process, the label information of source domain data is used to define $r$ in a supervised manner. In other words, based on the given information of $Y_s$, if two samples belong to the same emotion category, then $r=1$, otherwise $r=0$. A supervised $r$ can ensure the stability of the training process and the generalization ability of the model. The next key question is how to define a proper $g\left({x}_{i}^{s}, {x}_{j}^{s} ; \mathbf{\theta}\right)$ to compute the similarity between $x_i^s$ and $x_j^s$ in terms of the characterized interaction features (termed as $l_i$ and $l_j$). To make the similarity results locate in the range of $\left[0,1\right]$ and extract better and more robust feature representations for subsequent emotion recognition, we add a norm restriction on $l$ as

\begin{equation}
\label{Eq:pairwiseSourceLnorm}
        l^{norm}= \frac{l}{\left\|l\right\|_2}.
\end{equation}
The similarity of $l_{i}^{norm}$ and $l_{j}^{norm}$ is calculated as the cosine similarity, given as
\begin{equation}
    \label{Eq:pairwiseSourceG}
        g\left(x_{i}^{s}, x_{j}^{s} ; \theta\right)= l_{i}^{norm}\cdot l_{j}^{norm}=\frac{l_{i}^{s}\cdot l_{j}^{s}}{{\left\|l_{i}^{s}\right\|_2}{\left\|l_{j}^{s}\right\|_2}},
\end{equation}
where $\cdot$ refers to inner product operation. As stated in Chang \etal \cite{Chang2017}, the above-mentioned norm restriction can make the vector $l$ have a clustering function, and the elements in the vector represent the probability that the feature belongs to a certain category cluster. Overall, the objective function of pairwise learning on the source domain is defined as

\begin{equation}
\label{Eq:pairwiseSourceFinal}
    \begin{aligned}
        \mathcal{L}^{s}_{\text {pairwise}}(\theta)=
        \sum_{i, j} L\left(r_{i j}^{s}, \frac{l_{i}^{s}\cdot l_{j}^{s}}{{\left\|l_{i}^{s}\right\|_2}{\left\|l_{j}^{s}\right\|_2}}\right)
        +\beta\mathcal{R},
    \end{aligned}
\end{equation}
which is termed as \textbf{supervised pairwise loss} in Fig. \ref{fig:PR-PLmodel}. Here, $\theta=\left\{\theta_f,S\right\}$, $\theta_f$ is the parameter of feature extractor $f\left(\cdot\right)$, and $S$ is the defined bilinear transformation matrix for interaction feature extraction. Besides, to avoid redundant feature extraction, a soft regularization $\mathcal{R}$ is introduced with a weight parameter of $\beta$, which is defined as
\color{black}
\begin{equation}
    \label{Eq:pairwiseSourceR}
        \mathcal{R}=\left\|{P}^{T} {P}-{I}\right\|_{F}.
\end{equation}
Here, each row of the matrix $P$ refers to the prototype feature belonging to one emotion category, $\left\| \cdot \right\|_{F}$ is a $F$ norm of the matrix, and $I$ is an identity matrix.
\color{black}
The above loss function (Eq. \ref{Eq:pairwiseSourceFinal}) could be interpreted as a clustering loss, instead of emotion category classification loss. The main optimization goal is to gather the EEG samples that may belong to the same emotion category and separate the EEG samples that do not belong to the same category. The vector $l$ is the characterized interaction feature for mapping to a non-linear informative feature space by measuring the interaction relationship between the sample features and all the available prototype features.


\subsection{Pairwise learning on target domain}
\label{sec:pairwiseLearnTarget}
In this paper, besides of source domain, we also introduce the pairwise learning on the target domain to improve the feature separability in the target domain, as
\begin{equation}
\label{Eq:pairwiseTargetL}
        \mathcal{L}_{\text {pairwise}}^{t}\left(\theta_{f}, S\right)=
        \sum_{i, j} L\left(r_{i j}^{t}, \frac{l_{i}^{t}\cdot l_{j}^{t}}{{\left\|l_{i}^{t}\right\|_2}{\left\|l_{j}^{t}\right\|_2}}\right),
\end{equation}
which is termed as \textbf{unsupervised pairwise loss} in Fig. \ref{fig:PR-PLmodel}. Here, $l^t$ is the interaction features of the target domain data characterized in Eq. \ref{Eq:interactionFea}. The scalar $r_{ij}^{t}$ symbolizes the pairing relationship of the samples in the target domain. Since the label information of the target domain is completely missing in the training process, we cannot accurately obtain the pairing relationship as the source domain. To address this issue, we introduce an adaptive thresholding method to generate the valid pseudo labels and define the pairing relationship of the target domain data. Suppose that $r_{ij}^{t}$ is defined as 

\begin{equation}
\label{Eq:pairwiseTargetPseudo}
    \begin{aligned}
 r_{ij}^{t}:=\left\{\begin{array}{l}
    1, \text { if } {l_{i}^{t}} \cdot {l_{j}^{t}} \geq \tau_u \\
    0, \text { if } {l_{i}^{t}} \cdot {l_{j}^{t}}<\tau_l, \quad i, j=1, \cdots, n, \\
    \end{array}\right.
    \end{aligned}
\end{equation}
where $\tau_u$ and $\tau_l$ are the upper and lower bounds to select valid pairs with high confidence for unsupervised pairwise learning (valid pair selection). Here, if the calculated pairwise similarity is higher or equal to the defined upper bound ($\tau_u$), then the corresponding pseudo label {$r_{ij}^{t}$} would be assigned to 1; while if the calculated pairwise similarity is lower than the defined lower bound ($\tau_l$), then the corresponding pseudo label {$r_{ij}^{t}$} would be assigned to 0. For the other pairs that do not meet the threshold requirement, we would consider that the model is uncertain about whether the sample pair is paired or not and consider these pairs as invalid results. In order to prevent incorrect optimization, the invalid pairs would be temporarily excluded and will not participate in training and loss calculations at the current training round. 

In the early training stage, the classification performance in the target domain is not good enough. In order to ensure the training stability, we set a strict upper threshold ($\tau_u$) and a lower threshold ($\tau_l$) and exclude most of the pair results in the target domain. Along with the enhancement of model performance in the target domain, we can gradually lower the upper threshold and raise the lower threshold and allow more samples to participate in the training part. In other words, {along with the increase in training steps,} more pair samples in the target domain will be included for model learning. Here, we form a non-linear dynamic update for thresholding, as
\begin{equation}
\label{Eq:pairTargetUpperThreshold}
\tau_h^{t}=\tau_h^{t-1}-\frac{\tau_h^{t-1}-\tau_l^{t-1}}{maxepoch},
\end{equation}
\begin{equation}
\label{Eq:pairTargetLowerThreshold}
\tau_l^{t}=\tau_l^{t-1}+\frac{\tau_h^{t-1}-\tau_l^{t-1}}{maxepoch},
\end{equation}
where $\tau_h^t$ represents the upper threshold of the current training round $t$, {$\tau_l^t$} represents the current lower threshold, and $maxepoch$ is the maximum training round. Based on the given initial values $\tau_h^0$ and $\tau_l^0$, the calculations of {$\tau_h^t$} and {$\tau_l^t$} could be considered as non-linear changes with respect to the training rounds as
\begin{equation}
\label{Eq:pairTargetNonLinear}
\left\{\begin{array}{l}
\tau_h^{t}=\tau_h^{0}-\frac{\tau_h^{0}-\tau_l^{0}}{2}\times\left(1-\left(\frac{2}{maxepoch}\right)^{t}\right)\\
\tau_l^{t}=\tau_l^{0}+\frac{\tau_h^{0}-\tau_l^{0}}{2}\times\left(1-\left(\frac{2}{maxepoch}\right)^{t}\right).
\end{array}\right.
\end{equation}

In all, combining the domain adversarial loss (Eq. \ref{Eq:DANNfull}), the pairwise learning loss in the source domain (Eq. \ref{Eq:pairwiseSourceFinal}, and the pairwise learning loss in the target domain (Eq. \ref{Eq:pairwiseTargetL}), the final objective function of PR-PL could be given as follows:
\begin{equation}
\label{Eq:finalLoss}
        \min _{\theta_{f},S}\max _{\theta_{d}} \mathcal{L}_{\text {pairwise }}^{s}\left(\theta_{f}, S\right)+\gamma\mathcal{L}_{\text {pairwise}}^{t}\left(\theta_{f}, S\right)
        -\lambda \mathcal{L}_{\text {disc}}\left(\theta_{f}, \theta_{d}\right),
\end{equation}
where $\gamma$ is a hyperparameter to control the importance of the pairing loss of the target domain, given as $\gamma=\delta\times\frac{epoch}{maxepoch}$. $epoch$ is the current training round, and $maxepoch$ is the maximum training round. Empirically, $\delta$ is set to 2.

\section{Experimental Results}
\label{sec:experiment}

\subsection{Emotional Databases and Data Preprocessing}
To have a fair comparison with the state-of-the-art methods, we validate our proposed model on two well-known public databases: SEED \cite{zheng2015investigating} and SEED-IV \cite{zheng2018emotionmeter}. In SEED database \cite{zheng2015investigating}, a total of 15 subjects were invited. Each subject performed three sessions on different days and each session contained 15 trials. A total of three emotions were elicited (negative, neutral, and positive). In SEED-IV database \cite{zheng2018emotionmeter}, a total of 15 subjects participated in the experiment. For each subject, a total of 3 sessions were performed on different days and each session contained 24 trials. A total of four emotions were elicited ( happiness, sadness, fear, and neutral). The EEG signals of both SEED and SEED-IV databases were simultaneously collected using the 62-channel ESI Neuroscan system.


For EEG preprocessing, the data sampling rate was first downsampled to 200Hz, and the contaminated noises (e.g. EMG and EOG) were manually removed. Then, the data were filtered by a band-pass filter of 0.3 Hz to 50Hz. For each trial, the data was divided into a number of segments with a length of 1s. Based on the pre-defined five frequency bands: Delta (1-3 Hz), Theta (4-7 Hz), Alpha (8-13 Hz), Beta (14-30 Hz), and Gamma (31-50 Hz), the corresponding differential entropy (DE) features were extracted to represent the logarithm energy spectrum in a specific frequency band and total 310 features (5 frequency band $\times$ 62 channels) were obtained for one EEG segment. Then, all the features were smoothed with the linear dynamic system (LDS) method, which can utilize the time dependency of emotion changes and filter out emotion unrelated and noisy EEG components\cite{LDS2010}.

\subsection{Implementation Results}
In our experiments, the feature extractor $f$ and discriminator $d$ are both made up of multilayer perceptron (MLP) with the Relu activation function. All the parameters are randomly initialized from a uniform distribution. The bilinear operator matrix $S$ is also randomly initialized. In the model architecture, the feature extractor structure is designed as 310 (input layer)-64 (hidden layer 1)-Relu activation-64 (hidden layer 2)-Relu activation-64 (output feature layer). \textcolor{black}{The discriminator structure is designed as 64 (input layer)-64 (hidden layer 1)-Relu activation-dropout layer-64 (hidden layer 2)-1 (output layer)-Sigmoid activation.} The size of matrix $S$ given in Eq. \ref{Eq:bilinear} is $64 \times 64$. Besides, we adopt an RMSprop optimizer for network training, which shows a better performance than the other classic optimizers. The learning rate is set to 1e-3 and the mini-batch size for training is 96. To avoid overfitting problems, we use $L2$ regularizes (1e-5) in the networks. \textcolor{black}{The regularization coefficient $\beta$ in Eq. \ref{Eq:pairwiseSourceFinal} is 0.01. The balance parameter $\gamma$ for pairwise learning on the target domain in Eq. \ref{Eq:finalLoss} is controlled by a constant factor $\delta$ of 2. The threshold $\tau_h^0$ and $\tau_l^0$ are given to 0.9 and 0.5 respectively.} All the models are trained on an NVIDIA GeForce RTX 2080 GPU, with CUDA 10.0 using the Pytorch API.


\subsection{Experiment Protocols}
To fully evaluate the robustness and stability of the proposed model and compare it with the existing literature, we validate PR-PL using four different validation protocols. \textbf{(1) Cross-subject cross-session leave-one-subject-out cross-validation.} We evaluate the model with a strict cross-subject cross-session leave-one-subject-out to fully estimate the model robustness on the unknown subject(s) and session(s). One subject's all sessions data are used as the target and the remaining subjects' all sessions are used as the source. We repeat the training validation until each subject's all sessions are treated as the target for once. Due to the variants in individuals and sessions, this evaluation protocol poses a great challenge to the model's effectiveness in the EEG-based emotion recognition tasks. \textbf{(2) Cross-subject single-session leave-one-subject-out cross-validation.} It is the most widely used validation scheme in the EEG-based emotion recognition tasks \cite{Zheng2015CA,2018WGAN,li2019multisource,LiJDA2020}. One subject's one-session data is treated as the target and the other remaining subjects are used as the source. We repeat the training validation process until each subject is treated as the target for once. Same as the other studies, we only consider the first session in this type of cross-validation. \textbf{(3) Within-subject cross-session leave-one-session-out cross-validation.} Similar to the existing methods, a time-series cross-validation method is adopted here, where the past data is used to predict current or future data. For one subject, the first two sessions are used as the source, and the latter session is used as the target. The average accuracies and standard deviations across subjects are calculated as the final results. \textbf{(4) Within-subject single-session cross-validation.} Following the validation protocol presented in the existing studies \cite{zheng2015investigating,zheng2018emotionmeter}, for each session of one subject, we use the first 9 (SEED) or 16 (SEED-IV) trials as the source and the rest 6 (SEED) or 8 (SEED-IV) trials as the target. The results are reported as the average performance across all the subjects. In the following performance comparison across four different validation protocols, the model results reproduced by us are indicated by `*'.

\subsection{Cross-subject cross-session leave-one-subject-out cross-validation results}

To verify the model efficiency and stability on both cross-subject and cross-session conditions, we verify the proposed PR-PL using cross-subject cross-session leave-one-subject-out cross-validation on both SEED and SEED-IV databases. As reported in Table \ref{tab:seedfullycross} and Table \ref{tab:seedivfullycross}, the results show our proposed model achieves the highest results, where the emotion recognition performance of PR-PL is 85.56\%$\pm$4.78\% for three-class classification task on SEED and 74.92\%$\pm$7.92\% for four-class classification task on SEED-IV. Compared to the existing studies, the proposed PR-PL increases the classification accuracy to 3.39\% and 1.08\% for SEED and SEED-IV, with smaller standard deviations. These results demonstrate the proposed PR-PL has better affective effectiveness with higher recognition accuracy and better generalizability.

\begin{table}[]
\begin{center}
\caption{\textcolor{black}{The mean accuracies (\%) and standard deviations (\%) of emotion recognition on SEED database using cross-subject cross-session leave-one-subject-out cross-validation. Here, the model results reproduced by us are indicated by `*'.}}
\label{tab:seedfullycross}
\scalebox{1}{
\color{black}
\begin{tabular}{lc|lc}
\toprule
Methods   & $P_{acc}$    & Methods   & $P_{acc}$    \\ 
\midrule
\multicolumn{4}{c}{\textbf{\textit{Traditional machine learning methods}}} \\ 
\midrule
RF*\cite{Breiman2001RF}    & 69.60$\pm$07.64     & KNN*\cite{KNN1982}     &60.66$\pm$07.93  \\
SVM*\cite{SVM1999}         & 68.15$\pm$07.38     & Adaboost*\cite{2006Boost} &71.87$\pm$05.70  \\
TCA*\cite{TCA2010}         & 64.02$\pm$07.96     & CORAL*\cite{CORAL2016} &68.15$\pm$07.83  \\
SA*\cite{SA2013}           & 61.41$\pm$09.75     & GFK*\cite{GFK2012}     &66.02$\pm$07.59  \\ 

\midrule
\multicolumn{4}{c}{\textit{\textbf{Deep learning methods}}}\\ 
\midrule
DCORAL*  \cite{Dcoral2016}  & 81.97$\pm$05.16  & {DAN*\cite{longDAN2015} }         & 81.04$\pm$05.32  \\
DDC*     \cite{DDC2014}     & 82.17$\pm$04.96  & DANN* \cite{ganin2016domain} & 81.08$\pm$05.88  \\

\midrule
\multicolumn{3}{l}{\textbf{PR-PL}} & \textbf{85.56$\pm$04.78}\\
\bottomrule
\end{tabular}
}
\end{center}
\end{table}

\begin{table}[]
\begin{center}
\caption{\textcolor{black}{The mean accuracies (\%) and standard deviations (\%) of emotion recognition on SEED-IV database using cross-subject cross-session leave-one-subject-out cross-validation. Here, the model results reproduced by us are indicated by `*'.}}
\label{tab:seedivfullycross}
\scalebox{1}{
\color{black}
\begin{tabular}{lc|lc}
\toprule
Methods   & $P_{acc}$   & Methods   & $P_{acc}$    \\ 
\midrule
\multicolumn{4}{c}{\textbf{\textit{Traditional machine learning methods}}} \\ 
\midrule
RF*\cite{Breiman2001RF}    & 50.98$\pm$09.20  & KNN*\cite{KNN1982}  &40.83$\pm$07.28  \\
\textcolor{black}{SVM\cite{He2018DAN}}  & \textcolor{black}{51.78$\pm$12.85} &Adaboost*\cite{2006Boost}       &53.44$\pm$09.12\\
\textcolor{black}{TCA\cite{BiHDM2019}}     &56.56$\pm$13.77  & CORAL*\cite{CORAL2016} &49.44$\pm$09.09  \\
\textcolor{black}{SA\cite{BiHDM2019}}     & 64.44$\pm$09.46  & GFK*\cite{GFK2012}         & 45.89$\pm$08.27\\
\textcolor{black}{KPCA\cite{He2018DAN}}  & \textcolor{black}{51.76$\pm$12.89} &\textcolor{black}{DNN\cite{He2018DAN}}  & \textcolor{black}{49.35$\pm$09.74}\\
\midrule
\multicolumn{4}{c}{\textit{\textbf{Deep learning methods}}}\\ 
\midrule
DGCNN  \cite{song2018eeg}      & 52.82$\pm$09.23  & DAN \cite{He2018DAN}             & 58.87$\pm$08.13  \\
RGNN   \cite{zhong2020eeg}     & 73.84$\pm$08.02  & BiHDM \cite{BiHDM2019}           & 69.03$\pm$08.66  \\
BiDANN\cite{li2018bi}       & 65.59$\pm$10.39     & \textcolor{black}{DANN\cite{He2018DAN}} & \textcolor{black}{54.63$\pm$08.03}  \\
\midrule
\multicolumn{3}{l}{\textbf{PR-PL}} & \textbf{74.92$\pm$07.92}  \\
\bottomrule
\end{tabular}
}
\end{center}
\end{table}

\subsection{Cross-subject single-session leave-one-subject-out cross-validation results}
Table \ref{tab:seedCompare} summarizes the model results in cross-subject single-session leave-one-subject-out recognition task and compare the performance with the literature. All the results are presented in terms mean$\pm$standard deviation. The results show our proposed model (PR-PL) achieves the best performance (93.06\%), with a standard deviation of 5.12\%. Our PR-PL leads 2.14\% against the reported best results in the literature. Especially, compared to the latest proposed DANN based deep transfer learning networks (e.g. ATDD-DANN \cite{Du2020}, R2G-STNN\cite{li2019regional}, BiHDM\cite{BiHDM2019}, BiDANN\cite{li2018bi}, and WGAN-GP\cite{2018WGAN}), the proposed PR-PL with pairwise learning can avoid the inherent defects of DANN design (e.g. only considers feature separability on source domain) and well address the individual differences and noisy labeling issues in aBCI applications.

\begin{table}[]
\begin{center}
\caption{\textcolor{black}{The mean accuracies (\%) and standard deviations (\%) of emotion recognition on SEED database using cross-subject single-session leave-one-subject-out cross-validation. Here, the model results reproduced by us are indicated by `*'.}}
\label{tab:seedCompare}
\scalebox{1}{
\color{black}
\begin{tabular}{lc|lc}
\toprule
Methods   & $P_{acc}$   & Methods   & $P_{acc}$    \\ 
\midrule
\multicolumn{4}{c}{\textbf{\textit{Traditional machine learning methods}}} \\ 
\midrule
\textcolor{black}{TKL\cite{li2018bi}}      & 63.54$\pm$15.47 & \textcolor{black}{T-SVM\cite{li2018bi}}&72.53$\pm$14.00           \\
\textcolor{black}{TCA\cite{BiHDM2019}}      & 63.64$\pm$14.88 & \textcolor{black}{TPT\cite{He2018DAN}}& \textcolor{black}{75.17$\pm$12.83}           \\
\textcolor{black}{KPCA\cite{He2018DAN}}     & 61.28$\pm$14.62 & \textcolor{black}{GFK\cite{BiHDM2019}}       & 71.31$\pm$14.09           \\
\textcolor{black}{SA\cite{BiHDM2019}}       & 69.00$\pm$10.89  & \textcolor{black}{DICA\cite{Dresnet2019}}      & 69.40$\pm$07.80           \\ 
\textcolor{black}{DNN\cite{He2018DAN}}  & 61.01$\pm$12.38 & \textcolor{black}{SVM\cite{He2018DAN}}   & 58.18$\pm$13.85          \\
\midrule
\multicolumn{4}{c}{\textit{\textbf{Deep learning methods}}}\\ 
\midrule
DGCNN  \cite{song2018eeg}      & 79.95$\pm$09.02  & DAN \cite{He2018DAN}             & 83.81$\pm$08.56  \\
RGNN   \cite{zhong2020eeg}     & 85.30$\pm$06.72  & BiHDM \cite{BiHDM2019}           & 85.40$\pm$07.53  \\
WGAN-GP\cite{2018WGAN}         & 87.10$\pm$07.10  & \textcolor{black}{MMD\cite{LiJDA2020}}& 80.88$\pm$10.10\\
ATDD-DANN\cite{Du2020}         & 90.92$\pm$01.05  & JDA-Net\cite{LiJDA2020}          & 88.28$\pm$11.44  \\
R2G-STNN\cite{li2019regional}  & 84.16$\pm$07.63  & SimNet*\cite{Pinheiro2018}       & 81.58$\pm$05.11  \\
BiDANN\cite{li2018bi}          & 83.28$\pm$09.60  & DResNet\cite{Dresnet2019}        & 85.30$\pm$08.00\\
\textcolor{black}{ADA\cite{LiJDA2020}}              & 84.47$\pm$10.65  & DANN\cite{LiJDA2020}             & 81.65$\pm$09.92  \\
\midrule
\multicolumn{3}{l}{\textbf{PR-PL}} & \textbf{93.06$\pm$05.12}  \\
\bottomrule
\end{tabular}
}
\end{center}
\end{table}


\subsection{Within-subject cross-session cross-validation results}

By calculating the average and standard deviation of the experimental results of each subject, the final within-subject cross-session cross-validation results are reported in Table \ref{tab:seedsession3} for the SEED database and Table \ref{tab:seedivsession3} for the SEED-IV database. For both databases, our proposed PR-PL achieves the highest recognition performance compared with the state-of-the-art methods (including both traditional machine learning methods and deep learning methods), where the results are 93.18\%$\pm$6.55\% and 74.62\%$\pm$14.15\% for SEED (three-class emotion recognition) and SEED-IV (four-class emotion recognition), respectively. 

\begin{table}[]
\begin{center}
\color{black}
\caption{\textcolor{black}{The mean accuracies (\%) and standard deviations (\%) of emotion recognition on SEED database using within-subject cross-session cross-validation. Here, the model results reproduced by us are indicated by `*'.}}
\label{tab:seedsession3}
\scalebox{1}
{\color{black}
\begin{tabular}{lc|lc} 
\toprule
Methods & $P_{acc}$ & Methods  & $P_{acc}$   \\ 
\midrule
\multicolumn{4}{c}{\textbf{\textit{Traditional machine learning methods}}}      \\ 
\midrule
RF*\cite{Breiman2001RF}    & 76.42$\pm$11.15     & KNN*\cite{KNN1982}     &75.68$\pm$13.82  \\
TCA*\cite{TCA2010}         & 74.27$\pm$12.88     & CORAL*\cite{CORAL2016} &84.18$\pm$09.81  \\
SA*\cite{SA2013}           & 69.84$\pm$09.46     & GFK*\cite{GFK2012}     &78.79$\pm$09.39  \\ 
\midrule
\multicolumn{4}{c}{\textit{\textbf{\textbf{Deep learning methods}}}}            \\ 
\midrule
{DAN*\cite{longDAN2015} }       & 89.16$\pm$07.90    & SimNet*\cite{Pinheiro2018}  & 86.88$\pm$07.83 \\
DDC*\cite{DDC2014}          & 91.14$\pm$05.61    & \textcolor{black}{ADA\cite{LiJDA2020}}   & 89.13$\pm$07.13 \\
DANN*\cite{ganin2016domain} & 89.45$\pm$06.74    & \textcolor{black}{MMD\cite{LiJDA2020}}   & 84.38$\pm$12.05 \\
JDA-Net\cite{LiJDA2020}     & 91.17$\pm$08.11    & DCORAL*\cite{Dcoral2016} & 88.67$\pm$06.25 \\
\midrule
\multicolumn{3}{l}{\textbf{PR-PL}} & \textbf{93.18$\pm$06.55}  \\
\bottomrule
\end{tabular}
}
\end{center}
\end{table}

\begin{table}[]
\begin{center}
\color{black}
\caption{\textcolor{black}{The mean accuracies (\%) and standard deviations (\%) of emotion recognition on SEED-IV database using within-subject cross-session cross-validation. Here, the model results reproduced by us are indicated by `*'.}} 
\label{tab:seedivsession3}
\scalebox{1}
{\color{black}
\begin{tabular}{lc|lc}
\toprule
Methods & $P_{acc}$ & Methods  & $P_{acc}$   \\ 
\midrule
\multicolumn{4}{c}{\textbf{\textit{Traditional machine learning methods}}}      \\ 
\midrule
RF*\cite{Breiman2001RF}    & 60.27$\pm$16.36     & KNN*\cite{KNN1982}     &54.18$\pm$16.28  \\
TCA*\cite{TCA2010}         & 51.88$\pm$15.84     & CORAL*\cite{CORAL2016} &66.06$\pm$15.13  \\
SA*\cite{SA2013}           & 52.81$\pm$09.53     & GFK*\cite{GFK2012}     &56.14$\pm$12.15  \\ 
\midrule
\multicolumn{4}{c}{\textit{\textbf{\textbf{Deep learning methods}}}}            \\ 
\midrule
DCORAL  \cite{MEERNet2021}  & 65.10$\pm$13.20  & DAN \cite{MEERNet2021}        & 60.20$\pm$10.20  \\
DDC     \cite{MEERNet2021}     & 68.80$\pm$16.60  & MEERNet\cite{MEERNet2021} & 72.10$\pm$14.10  \\
\midrule
\multicolumn{3}{l}{\textbf{PR-PL}} & \textbf{74.62$\pm$14.15}  \\
\bottomrule
\end{tabular}
}
\end{center}
\end{table}

\subsection{Within-subject single-session cross-validation results}

Consistent with the evaluation method presented in the previous studies that only consider the first two sessions of the SEED database for experiments, we present the within-subject single-session results in Table \ref{tab:seed9v6}. It shows our proposed model obtains the best recognition performance of 94.84\%. Comparing the recognition results between cross-subject single-session (Table \ref{tab:seedCompare}) and within-subject single-session (Table \ref{tab:seed9v6}) emotion recognition tasks, the proposed PR-PL achieves the highest accuracies and at the same time perform the closest results on the two cross-validation methods (cross-subject single-session: 93.06$\pm$05.12; within-subject single-session: 94.84$\pm$09.16). For the other models, such as DGCNN\cite{song2018eeg}, BiDANN\cite{li2018bi}, R2G-STNN\cite{li2019regional}, RGNN\cite{zhong2020eeg}, and BiHDM\cite{BiHDM2019}, there exists a significant difference between cross-subject and within-subject results (9.09\% difference on average). This comparison demonstrates the efficiency and reliability of the proposed PR-PL in various emotion recognition applications. 

\begin{table}
\centering
\caption{ \textcolor{black}{The mean accuracies (\%) and standard deviations (\%) of emotion recognition on SEED database using within-subject single-session cross-validation. Here, the model results reproduced by us are indicated by `*'.}} 
\label{tab:seed9v6}
\color{black}
\begin{tabular}{lc|lc} 
\toprule
Methods   & $P_{acc}$   & Methods    & $P_{acc}$    \\ 
\midrule
\multicolumn{4}{c}{\textbf{\textit{Traditional machine learning methods}}}      \\ 
\midrule
\textcolor{black}{SVM\cite{BiHDM2019}}       & 83.99$\pm$09.72 & GRSLR\cite{GRSLR2019}     & 87.39$\pm$08.64  \\
\textcolor{black}{RF\cite{BiHDM2019}}  & 78.46$\pm$11.77 & GSCCA\cite{GSCCA2017}     & 82.96$\pm$09.95  \\
\textcolor{black}{CCA\cite{BiHDM2019}}      & 77.63$\pm$13.21 & DBN\cite{zheng2015investigating}& 86.08$\pm$08.34  \\
\midrule
\multicolumn{4}{c}{\textit{\textbf{\textbf{Deep learning methods}}}}            \\ 
\midrule
DGCNN\cite{2018SongDGCNN}      & 90.40$\pm$08.49 & RGNN\cite{zhong2020eeg}      & 94.24$\pm$05.95  \\
ATDD-DANN\cite{Du2020}         & 91.08$\pm$06.43 & BiHDM\cite{BiHDM2019}        & 93.12$\pm$06.06  \\
R2G-STNN\cite{li2019regional}  & 93.38$\pm$05.96 & SimNet*\cite{Pinheiro2018}    & 90.13$\pm$10.84  \\
BiDANN\cite{li2018bi}         & 92.38$\pm$07.04 & STRNN\cite{STRNN2019}        & 89.50$\pm$07.63  \\
\textcolor{black}{GCNN\cite{BiHDM2019}}   & 87.40$\pm$09.20 & {DANN\cite{BiHDM2019}}    & 91.36$\pm$08.30  \\ 
\midrule
\multicolumn{3}{l}{\textbf{PR-PL}}                      &\textbf{94.84$\pm$09.16}\\
\bottomrule
\end{tabular}
\end{table}

For the SEED-IV database, we calculate the performance on all three sessions as reported in the other studies and decode emotions into four categories (happiness, sadness, fear, and neutral). Our proposed model outperforms the existing studies, with the highest accuracy of 83.33\%, which leads to a 3.96\% increase as compared to the SOTA (79.37\%\cite{zhong2020eeg}).

\begin{table}
\centering
\caption{ \textcolor{black}{The mean accuracies (\%) and standard deviations (\%) of emotion recognition on SEED-IV database using within-subject single-session cross-validation. Here, the model results reproduced by us are indicated by `*'.}} 
\label{tab:seediv16v8}
\color{black}
\begin{tabular}{lc|lc} 
\toprule
Methods   & $P_{acc}$   & Methods    & $P_{acc}$    \\ 
\midrule
\multicolumn{4}{c}{\textbf{\textit{Traditional machine learning methods}}}      \\ 
\midrule
\textcolor{black}{SVM\cite{BiHDM2019}}  & 56.61$\pm$20.05 & GRSLR\cite{GRSLR2019}     & 69.32$\pm$19.57  \\
\textcolor{black}{RF\cite{BiHDM2019}}   & 50.97$\pm$16.22 & GSCCA\cite{GSCCA2017}     & 69.08$\pm$16.66  \\
\textcolor{black}{CCA\cite{BiHDM2019}}  & 54.47$\pm$18.48 & DBN\cite{zheng2015investigating}& 66.77$\pm$07.38 \\
\midrule
\multicolumn{4}{c}{\textit{\textbf{\textbf{Deep learning methods}}}}            \\ 
\midrule
DGCNN\cite{2018SongDGCNN}      & 69.88$\pm$16.29 & RGNN\cite{zhong2020eeg}      & 79.37$\pm$10.54  \\
\textcolor{black}{GCNN\cite{BiHDM2019}}& 68.34$\pm$15.42 & BiHDM\cite{BiHDM2019}        & 74.35$\pm$14.09  \\
\textcolor{black}{A-LSTM\cite{BiHDM2019}} & 69.50$\pm$15.45 & SimNet*\cite{Pinheiro2018}   & 71.38$\pm$13.12  \\
BiDANN\cite{li2018bi}         & 70.29$\pm$12.63 & {DANN\cite{BiHDM2019}}  & 63.07$\pm$12.66  \\
 
\midrule
\multicolumn{3}{l}{\textbf{PR-PL}}                      &\textbf{83.33$\pm$10.61}\\
\bottomrule
\end{tabular}
\end{table}

\section{Discussion and Conclusion} 
\label{sec:discussion}
To fully study the model performance, we evaluate the effect of different settings in PR-PL. Note that all the results presented in this {section} are based on the SEED database using the cross-subject single-session cross-validation evaluation protocol.

\subsection{Ablation Study}
We conduct the ablation study to systematically explore the effectiveness of different components in the proposed model and examine the corresponding contributions to the overall performance. As shown in Table \ref{tab:seedablation}, it is found that the introduction of domain adversarial training can greatly enhance the emotion recognition performance on the target domain. When the model is without discriminator and target domain information, the recognition accuracy reduces from 93.06\% to 83.30\%. Such a significant drop shows the significant impact of individual differences problem on model performance and highlights the great potential of transfer learning in aBCI applications. Besides, the results show a combination of pairwise learning on the source and target domain benefits to the model performance, where the recognition accuracy is increased by 6.35\% (from 87.5\% to 93.06\%). For the pseudo-labeling method, the corresponding accuracy increases from 89.92\% to 92.46\% when the pseudo-labeling method changes from fixed to linear dynamic. The accuracy further increases to 93.06\%, when a nonlinear dynamic-based adaptive pseudo-labeling method is adopted. The results show a non-linear dynamic pseudo-labeling could be helpful to screen out the valid paired samples and improve the model trainability. For the final loss function given in Eq. \ref{Eq:finalLoss}, instead of using a fixed weight for the pairwise loss on the target domain, we propose to update the weight gradually along with the training epochs to prevent model learning failures in the early training stage and balance the relationships among different losses. The benefit of a dynamic $\gamma$ is also reflected in the ablation study, where the recognition accuracy increases from 89.47\% and 93.06\%.

\begin{table}
\centering
\caption{\textcolor{black}{The ablation study of our proposed model.}}
\color{black}
\label{tab:seedablation}
\scalebox{1}{
\begin{tabular}{lccc} 
\toprule
\multicolumn{3}{l}{\textbf{\textit{Ablation study about training strategy}}}& $P_{acc}$ \\
\midrule
\multicolumn{3}{l}{w/o discriminator and target information}&83.30$\pm$04.21  \\
\multicolumn{3}{l}{w/o pairwise learning on the source and target} &87.50$\pm$06.64  \\
\multicolumn{3}{l}{w/o pairwise learning on the target} &88.81$\pm$06.63  \\
\multicolumn{3}{l}{\textcolor{black}{w/o prototypical representation}}           &\textcolor{black}{91.00$\pm$04.65}  \\
\multicolumn{3}{l}{w/o thresholding for pseudo label generation}           &92.13$\pm$05.90  \\

\midrule
\multicolumn{3}{l}{\textbf{\textit{About hyperparameter controlling strategy}}}& $P_{acc}$ \\
\midrule
\multicolumn{3}{l}{w/ fixed pseudo-labeling}         &89.92$\pm$07.21  \\
\multicolumn{3}{l}{w/ linear dynamic pseudo-labeling}&92.46$\pm$04.95  \\
\multicolumn{3}{l}{w/ fixed $\gamma$ for target pairwise loss}&89.47$\pm$10.22 \\
\midrule
\multicolumn{3}{l}{\textbf{\textbf{PR-PL}}}                      
&\textbf{93.06$\pm$05.12} \\
\bottomrule
\end{tabular}
}
\end{table}

\subsection{Effect of Noisy Labels}
To further verify the model robustness during noisy label learning, we randomly contaminate the source labels with $\eta$\% noises and test the corresponding model performance on unknown target data. Specifically, we replace $\eta$\% real labels in $Y_s$ using randomly generated labels and train the model in supervised learning. Then, we test the trained model performance on the target domain. Note here that the noisy contamination is only conducted on the source domain, as the target domain needs to be used for model evaluation. In the implementation, $\eta$\% value is adjusted to 10\%, 20\%, and 30\%, respectively. The corresponding model accuracies with the standard deviations are 89.22\%$\pm$6.05\%, 88.39\%$\pm$6.73\%, and 87.71\%$\pm$5.02\%. It shows that, with an increase of label noise ratio from 10\% to 30\%, the model performance decreases slightly, with a decrease rate of 1.69\%. These results demonstrate the proposed PR-PL is a reliable model which has a higher tolerance to noisy labels. {In the future works, the recently proposed novel methods, such as \cite{xiao2019discriminative} and \cite{jin2021robust}, could be incorporated to further eliminate more general noises in EEG signals and improve the model stability in the cross-corpus applications.}

\subsection{Confusion Matrices}
\begin{figure*}[h]
\begin{center}
\includegraphics[width=1\textwidth]{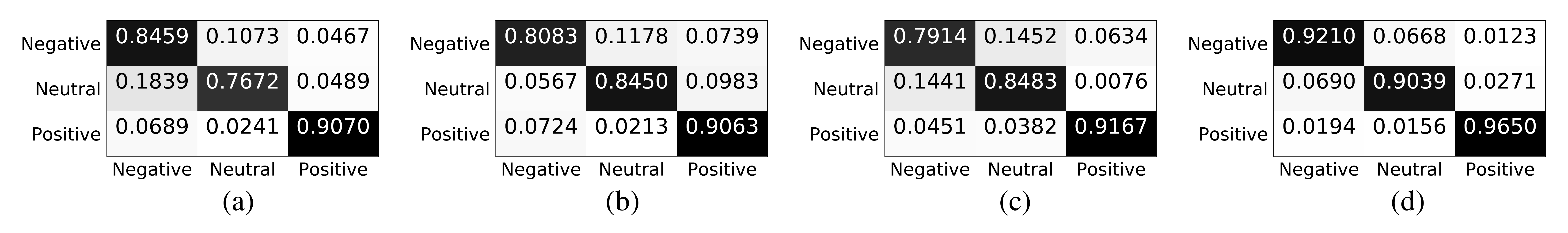}
\caption{\textcolor{black}{Confusion matrices of different models. (a) BiDANN \cite{li2018bi}, (b) BiHDM \cite{BiHDM2019}, (c) RGNN \cite{zhong2020eeg}, and (d) PR-PL.}}
\label{fig:confusion_matrix}
\end{center}
\end{figure*}

To qualitatively study the model performance in each emotion class, we visually analyze the confusion matrices and compare results with the latest models \cite{li2018bi,BiHDM2019,zhong2020eeg}. As shown in Fig.\ref{fig:confusion_matrix}, it shows all the models are good at distinguishing positive emotions from other emotions (the recognition rates are all above 90\%), but it is relatively poor at distinguishing negative and neutral emotions. For example, the recognition rate of neural in BiDANN \cite{li2018bi} is even lower than 80\% (76.72\%). Compared to the existing methods ((a), (b), and (c)), our proposed model can enhance the model recognition ability, especially for distinguishing neutral and negative emotions. As shown in (d), the recognition rates for negative, neutral, and positive emotions are 92.10\%, 90.39\%, and 96.50\%. 

\subsection{Visualization of Learned Representation}
\begin{figure}[h]
\begin{center}
\includegraphics[width=0.5\textwidth]{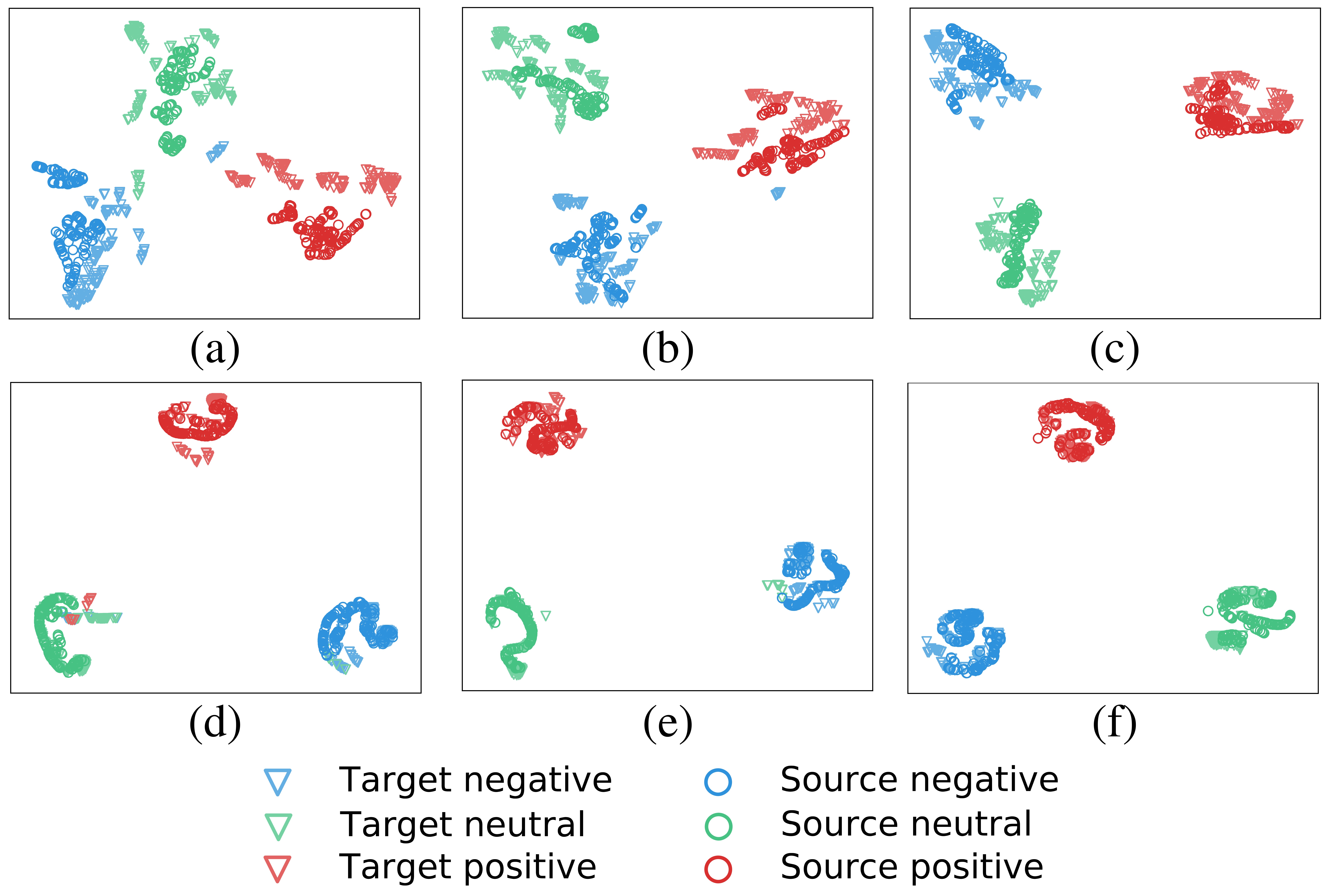}
\caption{\textcolor{black}{T-SNE visualization of the learned features from the source and the target domains using different model settings. (a), (b) and (c) are the sample features extracted by w/o pairwise learning on the source and target, w/o pairwise learning on the target, and PR-PL. (d), (e) and (f) are the interaction features extracted by w/o pairwise learning on the source and target, w/o pairwise learning on the target, and PR-PL.}}
\label{fig:tsne}
\end{center}
\end{figure}

To verify the effectiveness of the proposed model from a more intuitive perspective, we visualize the characterized sample and interaction features of source and target domains using T-SNE\cite{2008Tsne} in Fig. \ref{fig:tsne}. Here, we randomly select 500 samples from the source and 500 samples from the target for visualization of the learned feature representation. Compared to the representation learned by the other model settings (w/o pairwise learning on the source and target and w/o pairwise learning on the target), the representation learned by the proposed PR-PL forms more separated emotional clusters. Comparing the extracted sample features (c) and interaction features (f) by the proposed PR-PL, the separability of the extracted interaction features from different emotion classes is further enlarged and at the same time, the concentration of the feature distribution for each emotion is also improved.

\subsection{Conclusion}
The paper proposes a novel transfer learning framework with prototypical representation-based pairwise learning (PR-PL), that characterizes EEG data with prototypical representations and formulates the EEG-based emotion recognition task as pairwise learning. We evaluate our proposed model on two well-known emotional databases (SEED and SEED-IV) under four cross-validation protocols (cross-subject single-session, within-subject single-session, within-subject cross-session, and cross-subject cross-session) and compare it with the existing state-of-the-art methods. Our extensive experimental results show PR-PL achieves the best results on all four cross-validation protocols and demonstrate the advantage of PR-PL in tackling individual differences and noisy labeling issues in aBCI systems.

\section{Conflicts of Interest}
The authors declare that they have no conflicts of interest.

\section{Acknowledgment}
This work was supported in part by the National Natural Science Foundation of China under Grant 61906122, in part by Shenzhen-Hong Kong Institute of Brain Science-Shenzhen Fundamental Research Institutions (2021SHIBS0003), in part by the Tencent “Rhinoceros Birds”-Scientific Research Foundation for Young Teachers of Shenzhen University, and in part by the High Level University Construction under Grant 000002110133.

\ifCLASSOPTIONcaptionsoff
  \newpage
\fi

\bibliographystyle{IEEEtran}
\bibliography{references}

\end{document}